\documentclass[twocolumn,showpacs,preprintnumbers,amsmath,amssymb]{revtex4}

\usepackage{bm}
\usepackage[colorlinks=true,linkcolor=blue,citecolor=blue,urlcolor=blue]{hyperref}
\usepackage{color}

\usepackage{graphicx}
\usepackage{epstopdf}

\begin{document}


\title{Tunable exchange bias effect in magnetic  Bi$_{0.9}$Gd$_{0.1}$Fe$_{0.9}$Ti$_{0.1}$O$_3$ nanoparticles at temperatures up to 250 K}

\author{M. A. Basith}
\email[Author to whom correspondence should be addressed (e-mail): ]{mabasith@phy.buet.ac.bd}
\author{F. A. Khan}
\affiliation{Department of Physics, Bangladesh University of Engineering and Technology, Dhaka-1000, Bangladesh.
}
\author{Bashir Ahmmad}
\email[Author to whom correspondence should be addressed (e-mail): ]{arima@yz.yamagata-u.ac.jp}
\affiliation{Graduate School of Science and Engineering, Yamagata University, 4-3-16 Jonan, Yonezawa 992-8510, Japan.}
\author{Shigeru Kubota, Fumihiko Hirose}
\affiliation{Graduate School of Science and Engineering, Yamagata University, 4-3-16 Jonan, Yonezawa 992-8510, Japan.}
\author{D.-T. Ngo}
\email[Author to whom correspondence should be addressed (e-mail): ]{dngo@nanotech.dtu.dk}
\affiliation{Department of Micro-and Nanotechnology, Technical University of Denmark, Kgs. Lyngby 2800, Denmark.}
\author{Q.-H. Tran, K. M{\o}lhave}
\affiliation{Department of Micro-and Nanotechnology, Technical University of Denmark, Kgs. Lyngby 2800, Denmark.}




\date{\today}

\begin{abstract}
The exchange bias (EB) effect has been observed in magnetic Bi$_{0.9}$Gd$_{0.1}$Fe$_{0.9}$Ti$_{0.1}$O$_3$ nanoparticles. The influence of magnetic field cooling on the exchange bias effect has also been investigated. The magnitude of the exchange bias field ($H_{EB}$) increases with the cooling magnetic field, showing that  the strength of the exchange bias effect is tunable by the field cooling. The $H_{EB}$ values are also found to be dependent on the temperature. This magnetically tunable exchange bias obtained at temperatures up to 250 K in Bi$_{0.9}$Gd$_{0.1}$Fe$_{0.9}$Ti$_{0.1}$O$_3$ nanoparticles may be worthwhile for potential applications. 
  
\end{abstract}

\maketitle
\section{Introduction} \label{I}
The exchange bias (EB) effect was discovered in 1956 by Meiklejohn and Bean when studying Co particles embedded in their native antiferromagnetic oxide (CoO) \cite{ref1}. This effect is widely used in different magnetic sensors and read heads for spintronic applications \cite{ref3}. The phenomenon manifests itself by a shift of the magnetization vs magnetic field ($M-H$) hysteresis loop of a magnetic system along the field axis when the system is cooled down through the N\'eel temperature of the antiferromagnetic component of the system. So far most of the experimental studies in this field have been focused on the investigations of specially prepared systems, such as core-shell nanoparticles with the ferromagnetic (FM) core coupled to the antiferromagnetic (AFM) shell \cite{ref1,ref2}, or thin films composed of FM and AFM thin layers \cite{ref3} or bi-layer systems like Co/CuMn where a spin glass like behavior has been  used to interpret the EB effect \cite{ref58}. The EB effect was observed not only in specially prepared systems but also in bulk perovskite manganites \cite{ref7,ref10} and cobaltites \cite{ref15} with spontaneous magnetic phase segregation. Interestingly, the EB effect attributed to spontaneous phase separation has recently been observed in a couple of multiferroic materials \cite{ref26}. However, almost all of the currently known materials with multifunctional activities demonstrate a low magnetic-ordering temperature which is in contrast to a room temperature ferromagnetic-transition temperature ($T_C$ $>$ 350 K) \cite{ref20} typically needed in applications. This is a clear obstacle to the exploitation of multiferroics in real applications at room temperature. It is worth noting that among the limited available multiferroics, BiFeO$_3$ stands out as the only multiferroic compound which is so far known to show both electric and magnetic ordering in a single phase well above the room temperature \cite{ref22}. The high characteristic phase transition temperatures, e.g., ferroelectric Curie temperature $T_C$ = 1100 K and antiferromagnetic  N\'eel temperature $T_N$ = 640 K \cite{ref20,ref25} of BiFeO$_3$ are promising for applications in magnetic and ferroelectric devices. The co-existence of multiferroicity and EB in structurally single phase BiFeO$_3$ is remarkable and needs to be explored to search for new multiferroic materials having technological applications. Recently, to ascertain the presence of EB effect in Na doped BiFeO$_3$ nanoparticle system, the training effect was studied \cite{ref24}.  For the 3 $\%$  Na doped BiFeO$_3$ nanoparticles, a monotonic decrease in the EB field have been observed with increase in the field cycles number (n). The experimentally observed training effect data points were found in accordance with Binek's model based on the antiferromagnetic (AFM) and ferromagnetic (FM) interface. Notably, the EB effect was observed in BiFeO$_3$ without any magnetic-field-annealing process through  $T_N$ \cite{ref22} (the conventional method of inducing unidirectional anisotropy \cite{ref27}). It has also been observed without using any alloy layers  \cite{ref28}, however, then the biasing strength of BiFeO$_3$ is observed to be very weak with $H_{EB}$ = 36 Oe \cite{ref28} at room temperature. Such a weak biasing effect is also observed at room temperature in a number of previously studied materials \cite{ref85}. Moreover, the effect is found to vanish above 100 K \cite{ref71} in some other materials, thus making such systems less attractive for applications.\\
In the present investigation, we have observed increasing exchange bias effect as temperatures drop, reaching 103 Oe only 50 K below room temperature in magnetic Bi$_{0.9}$Gd$_{0.1}$Fe$_{0.9}$Ti$_{0.1}$O$_3$ nanoparticles. The effect appears as an asymmetric shift of  the hysteresis loop towards the magnetic field axis both in zero field cooling (ZFC) and field cooling (FC) conditions. The exchange bias field ($H_{EB}$) was found to increase significantly as the cooling magnetic field ($H_{cool}$) was increased.

\section{Experimental details} \label{II}
\begin{figure}[!h]
\centering
\includegraphics[width=8cm]{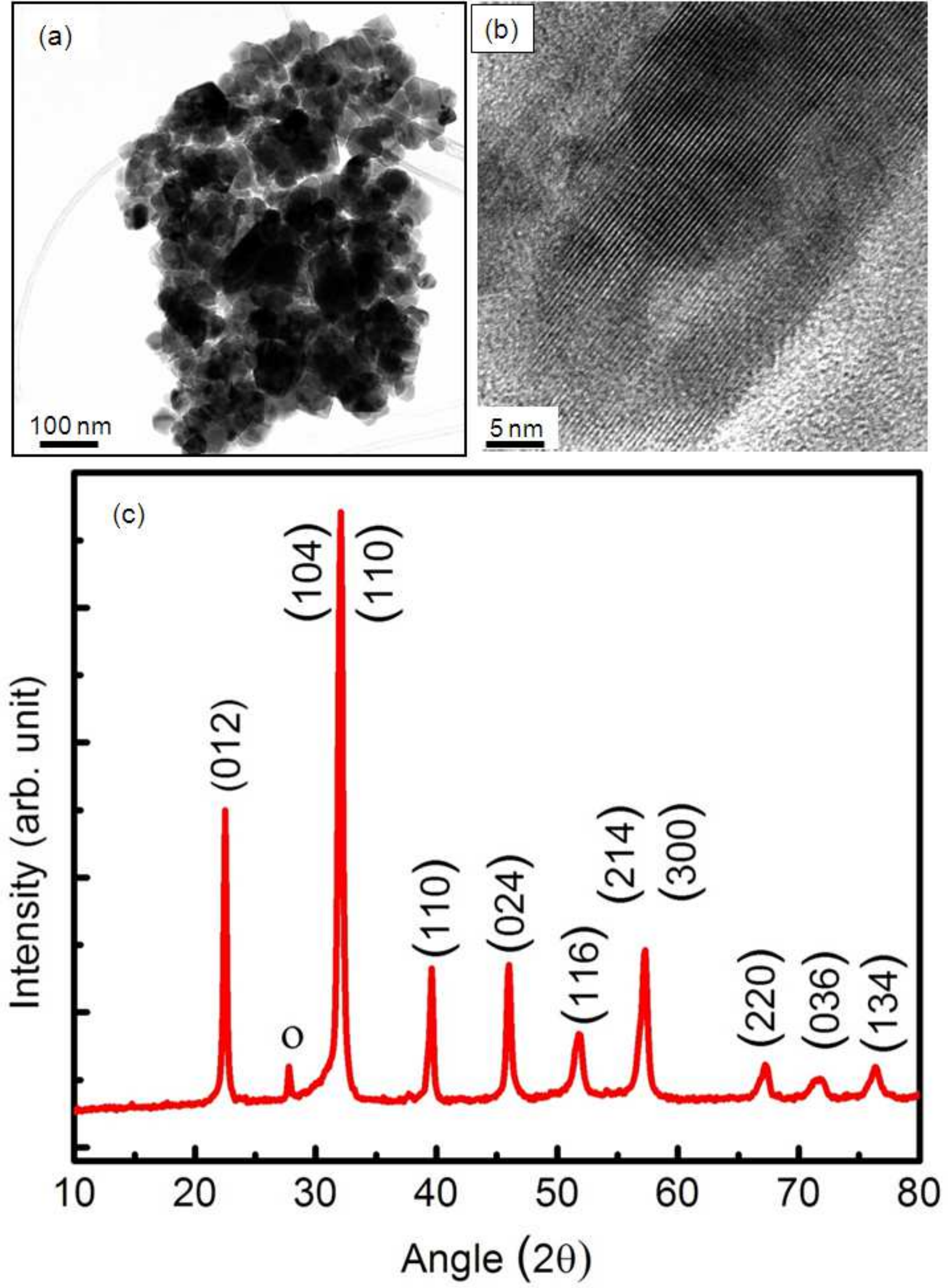}
\caption{(a) Bright field TEM image of Bi$_{0.9}$Gd$_{0.1}$Fe$_{0.9}$Ti$_{0.1}$O$_3$ nanoparticles prepared by ultrasonication technique \cite{ref37}. (b) HR TEM image showing the crystal plane of a monocrystalline nanoparticle. (c) X-ray diffraction patterns of Bi$_{0.9}$Gd$_{0.1}$Fe$_{0.9}$Ti$_{0.1}$O$_3$ nanoparticles  (The open circle denotes secondary phase). } \label{fig1}
\end{figure}

The mutiferroic ceramic with nominal composition  Bi$_{0.9}$Gd$_{0.1}$Fe$_{0.9}$Ti$_{0.1}$O$_3$  was prepared initially by conventional solid state reaction technique which was described in details  elsewhere \cite{ref36}. The ceramic pellets  were  ground again  into powder by manual grinding.  The Bi$_{0.9}$Gd$_{0.1}$Fe$_{0.9}$Ti$_{0.1}$O$_3$ nanoparticles were prepared directly from this bulk powder by using the sonication technique described in Ref. \cite{ref37}. The bulk powders were mixed with isopropanol with a ratio of 50 mg powder and 10 ml isopropanol and then put into an ultrasonic bath and was sonicated for 60 minutes. After around four hours, $\sim$ 35 \% of the mass was collected as  supernatant and was used for structural and magnetic characterization. The particle size was studied using transmission electron microscopy (TEM) imaging  that confirmed the formation of a large fraction of single-crystalline nanoparticles with a mean size of 40-100 nm as shown in figure \ref{fig1}(a). A high magnification TEM bright field image and the corresponding particle size distribution histogram deduced from a number of TEM images were shown in the supplemental figure 1(S1) \cite{ref21}. The high resolution (HR) TEM image, figure \ref{fig1}(b) shows the crytalline plane of the monocrystalline particle. The X-ray diffraction (XRD) patterns of the Bi$_{0.9}$Gd$_{0.1}$Fe$_{0.9}$Ti$_{0.1}$O$_3$ nanoparticles with rhombohedral crystal structure are shown in figure \ref{fig1}(c). In the synthesized nanoparticles, there is only one secondary phase  and it is labeled by an open circle, figure \ref{fig1}(c) compared to that of a significant number of different impurity peaks appeared in the corresponding bulk ceramic \cite{ref36}. Magnetization measurements  of magnetic Bi$_{0.9}$Gd$_{0.1}$Fe$_{0.9}$Ti$_{0.1}$O$_3$ nanoparticles were carried out using a Superconducting Quantum Interference Device (SQUID) Magnetometer (Quantum Design MPMS-XL7, USA) both at zero field cooling (ZFC) and field cooling (FC) processes. 

\section{Results and discussions} \label{III}
\begin{figure}[!h]
\centering
\includegraphics[width=8.5 cm]{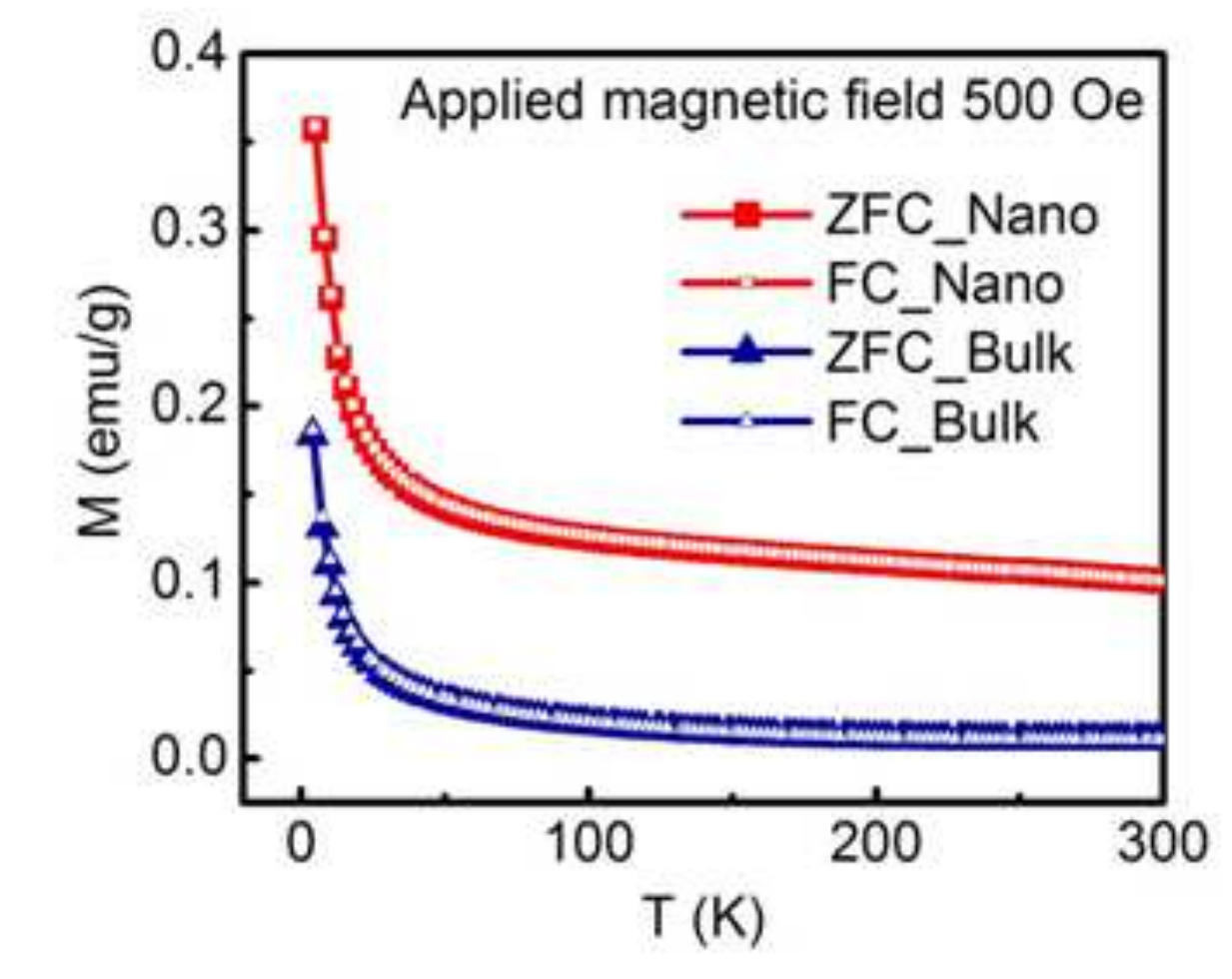}
\caption{Temperature dependence of magnetization ($M-T$ curves) of Bi$_{0.9}$Gd$_{0.1}$Fe$_{0.9}$Ti$_{0.1}$O$_3$ nanoparticles and bulk ceramic powders measured in ZFC and FC processes.} \label{fig2}
\end{figure}

First of all, to ensure the  difference in magnetization between synthesized nanoparticles and the corresponding bulk ceramic powders a temperature dependent magnetization measurement was carried out. Figure \ref{fig2} shows the temperature dependence of the magnetization ($M-T$) of Bi$_{0.9}$Gd$_{0.1}$Fe$_{0.9}$Ti$_{0.1}$O$_3$ nanoparticles and bulk ceramic materials in ZFC and FC processes. To perform the experiment the sample was cooled in the absence of the magnetic field (ZFC) to a low temperature. At the lowest achievable temperature a small magnetic field of 500 Oe is applied on the sample. Then the magnetization was measured as a function of temperature as the sample was heated back to room temperature in the 500 Oe field. The magnetization of nanoparticles is significantly higher (around ten times) over a wide range of temperature compared to that of bulk material. This enhancement is due to the reduced size of the phase pure nanoparticles \cite{ref37,ref38,ref72}. With decreasing particle size, the surface-to-volume ratio increases and therefore the contribution of the surface spins to the total magnetic moment of the particle increases \cite{ref72,ref77}. Both ZFC and FC $M-T$ curves of Bi$_{0.9}$Gd$_{0.1}$Fe$_{0.9}$Ti$_{0.1}$O$_3$ nanoparticles have not shown any bifurcation which indicates the absence of any spin flipping effect \cite{ref38, ref35,ref99}. The ZFC and FC curves were also found to coincide with each other in similar multiferroic materials \cite{ref76, ref78}. 

\begin{figure}[!h]
\centering
\includegraphics[width=8.5 cm]{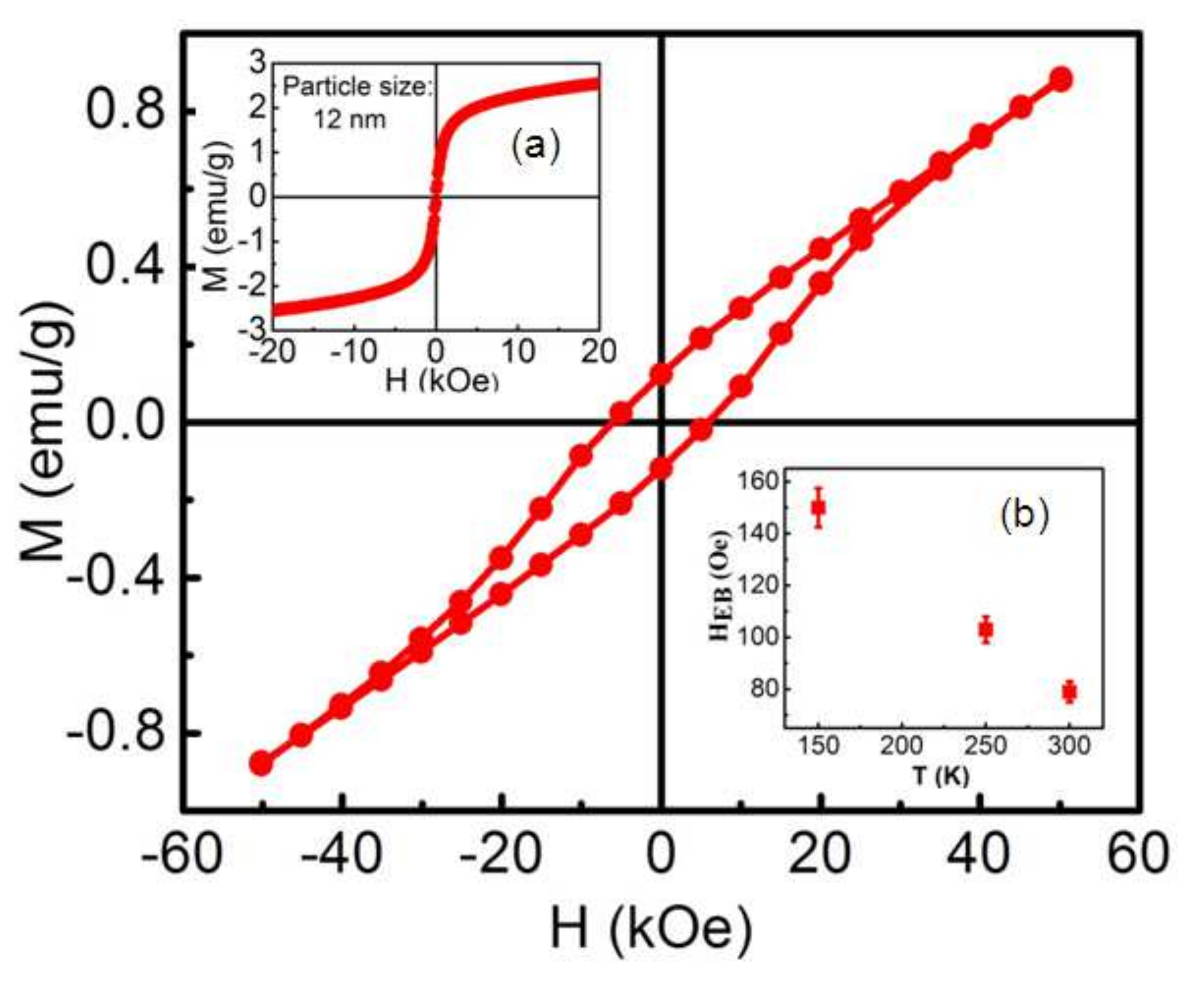}
\caption {The room temperature $M-H$ hysteresis loops of magnetic Bi$_{0.9}$Gd$_{0.1}$Fe$_{0.9}$Ti$_{0.1}$O$_3$ nanoparticles having average particle size 40-100 nm. The inset (a) shows the M-H loop of the nanoparticles having average size 12 nm \cite{ref37}. The inset (b) shows the variation of $H_{EB}$ as a function of temperature.} \label{fig3}
\end{figure}

An M-H hysteresis loop of the synthesized nanoparticles was carried out at room temperature, figure \ref{fig3}. The unsaturated hysteresis loop with an applied magnetic field of up to 50 kOe along with a large value of coercivity (around 6 kOe) clearly indicates the presence of both ferri/ferro and antiferromagnetic domains in these nanoparticles. In our previous investigation, for nanoparticles having 12 nm particle size we have observed a well-defined ferromagnetic hysteresis loop at room temperature with a small value of coercivity (53 Oe) \cite{ref37} as shown in the inset of figure \ref{fig3} (Inset (a)). Therefore, the presence of multiple magnetic domains in nanoparticles having particle sizes 40-100 nm provides the pre-condition to explore EB effect at/near room temperature.

This motivated us to study the possible exchange bias effect in  Bi$_{0.9}$Gd$_{0.1}$Fe$_{0.9}$Ti$_{0.1}$O$_3$ nanoparticles and therefore the $M-H$ hysteresis loops were carried out. The hystereis loop was taken at 300 K and then in separate experiments at 250 K and 150 K by cooling down the sample from 300 K. During the cooling process from 300 K to 250 K and 150 K we did not apply any magnetic field i.e. the experiments were carried out in ZFC condition. The room temperature $M-H$ hysteresis loop as well as the loops taken at 150 K and 250 K in zero field condition exhibit an asymmetric shift towards the magnetic field axes \cite{ref41}. This is a signature of the presence of an exchange bias effect in Bi$_{0.9}$Gd$_{0.1}$Fe$_{0.9}$Ti$_{0.1}$O$_3$ nanoparticles \cite{ref41}. The exchange bias field from the loop asymmetry along the field axis can be quantified as $H_{EB} = -(H_{c1}+H_{c2})/2$ where $H_{c1}$ and $H_{c2}$ are the left and right coercive fields, respectively \cite{ref7,ref41}. The variation of $H_{EB}$ as a function of temperature is shown in the inset of figure \ref{fig3} (b). Notably, at 300 K, for nanoparticles having particle size 40-100 nm, the exchange bias field is 79 Oe which is significantly higher than the value reported in Ref.(\cite{ref28}, $H_{EB}$ = 36 Oe) for bulk BiFeO$_3$ ceramic. In the case of our previously investigated nanoparticles having 12 nm particle size \cite{ref37}, their nearly saturated hysteresis loop demonstrated a ferromagnetic nature of the synthesized nanoparticles and their magnetization values were much larger than those of nanoparticles having 40-100 nm size and also than those of their bulk counterparts. But, as the system having 12 nm particle size is turned into almost one single ferromagnetic domain, any notable exchange effect was absent. The effect of increased particle size on the magnetic behavior and in particular on the exchange bias effect of this kind of multiferroic system (undoped BiFeO$_3$) was reported in Ref. \cite{ref72}. The exchange bias field was found to decrease with reduced particle size which have comparatively higher saturation magnetization.  For 14 nm BiFeO$_3$ nanoparticles, the hysteresis loop was fairly saturated and the biasing field was insignificant i.e. only 2.5 Oe \cite{ref72}.
	
As there is a substantial value of $H_{EB}$ at 150 K and 250 K in ZFC condition, therefore, in the subsequent experiments, the magnetic hysteresis loops were carried out at these two temperatures by cooling down the sample from 300 K in various cooling fields ($H_{cool}$) ranging from 20 kOe to 60 kOe. In each experiment, the measuring magnetic fields were from -30 kOe to 30 kOe. Notably, using the SQUID magnetometer (Quantum Design MPMS-XL7, USA) it was not possible to heat the sample above 300 K, therefore the field cooling experiments were not possible to conduct at 300 K. 
\begin{figure}[!h]
\centering
\includegraphics[width=8.5 cm]{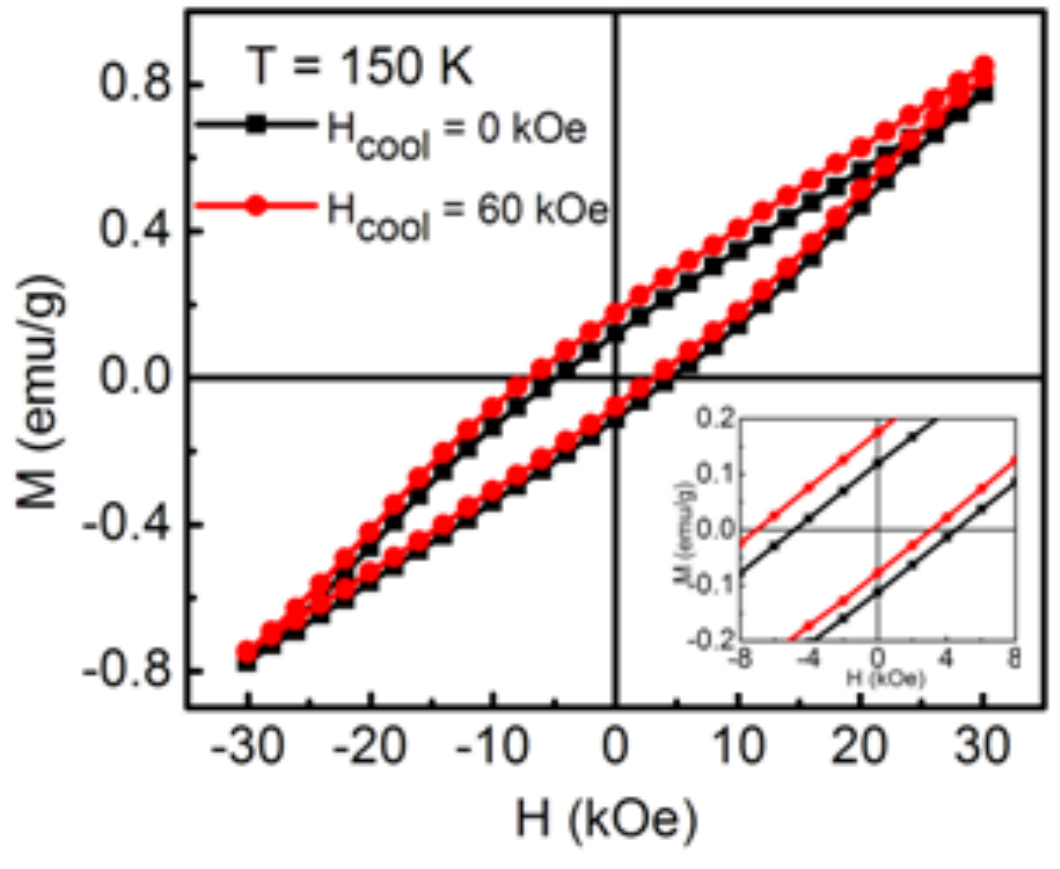}
\caption{$M-H$ hysteresis loop of  Bi$_{0.9}$Gd$_{0.1}$Fe$_{0.9}$Ti$_{0.1}$O$_3$ nanoparticles at 150 K. Inset: an enlarged view of $M-H$ hysteresis loops showing the asymmetric shifting of field axis of  Bi$_{0.9}$Gd$_{0.1}$Fe$_{0.9}$Ti$_{0.1}$O$_3$ nanoparticles for a cooling field of 60 kOe.} \label{fig4}
\end{figure}

The $M-H$ loops of Bi$_{0.9}$Gd$_{0.1}$Fe$_{0.9}$Ti$_{0.1}$O$_3$ nanoparticles at 150K after being cooled from 300 K in zero field and then in a separate experiment in 60 kOe fields are presented in figure \ref{fig4} as a typical example. Strong magnetic anisotropy in the material can be confirmed via very large coercive field (around 5.1 kOe) and unsaturated loops when applying high fields. The inset of figure \ref{fig4} shows an enlarged view of the central region of the loops of nanoparticles which demonstrate clearly  asymmetry of the loops along the field axis (inset of figure \ref{fig4}). More details of the loop asymmetry at different cooling fields and temperatures can be found in the supplemental figures 2(S2) and 3(S3) \cite{ref21}. The exchange bias field $H_{EB}$ values were calculated from the loop asymmetry at different temperatures for different cooling fields, $H_{cool}$. Along with $H_{EB}$, the coercive field ($H_c$) is also quantified by $H_c = (H_{c1}-H_{c2})/2$ \cite{ref42}. The $H_c$ values for nanoparticles at 150 K and 250 K are presented in supplemental table 1 \cite{ref21}. At high temperature (250 K), the highest coercivity is 5.1 kOe for a cooling magnetic fields of 60 kOe. The coercivity was found to increase with cooling magnetic fields as was also observed in multiferroic \cite{ref26} as well as in manganite system \cite{ref7}. 

\begin{figure}[!h]
\centering
\includegraphics[width=8.5cm]{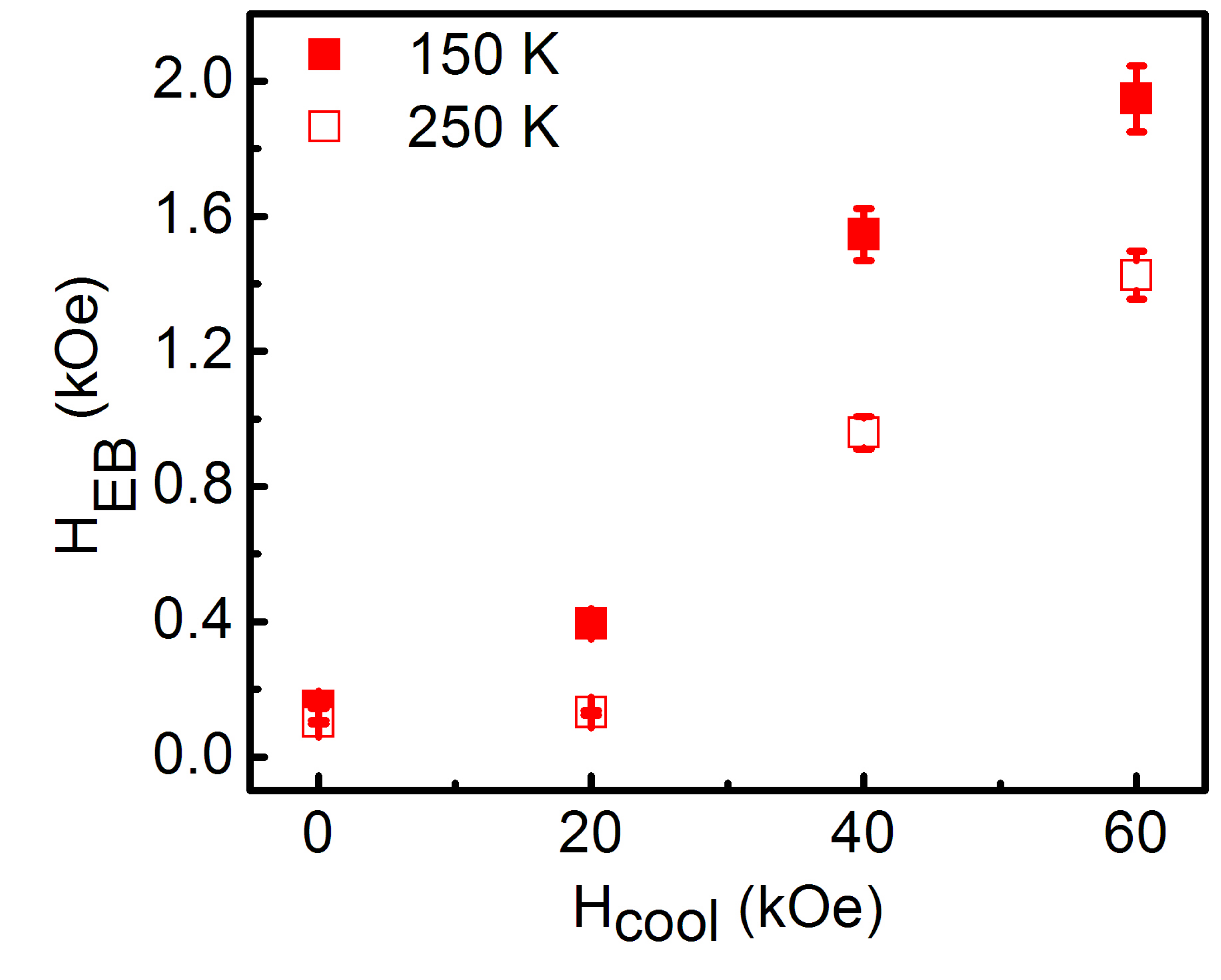}
\caption{Cooling field dependence of exchange bias field ($H_{EB}$) at 150 k and 250 K for Bi$_{0.9}$Gd$_{0.1}$Fe$_{0.9}$Ti$_{0.1}$O$_3$ nanoparticles.} \label{fig5}
\end{figure}
Figure \ref{fig5} illustrates the exchange bias effect at 150 k and 250 K in Bi$_{0.9}$Gd$_{0.1}$Fe$_{0.9}$Ti$_{0.1}$O$_3$ nanoparticles in which the $H_{EB}$ can be controlled by cooling field up to 60 kOe. The $H_{EB}$  values are reduced with increasing temperature (figure \ref{fig5} and figure \ref{fig3} (b)) which is similar to results reported in previous investigations \cite{ref26,ref101,ref71}.   It is worth mentioning that such a high magnitude of exchange bias field $H_{EB}$ at such high temperatures has never been observed so far in related materials system \cite{ref10,ref26} to the best of our knowledge. The $H_{EB}$  values as a function of cooling magnetic fields were also significantly higher in nanocrystalline La$_{1/3}$Sr$_{2/3}$FeO$_{3-\delta}$ \cite{ref97} compared to other reported values in structurally single phase alloys and compounds \cite{ref98}. However, the $H_{EB}$  values in  nanocrystalline La$_{1/3}$Sr$_{2/3}$FeO$_{3-\delta}$ were observed at 5 K \cite{ref97} whereas in this investigation the values were reported up to 250 K.  

As shown in figure \ref{fig5}, increasing cooling field, $H_{cool}$, gives rise to increase in $H_{EB}$. This observation ultimately indicates that the cooling magnetic field, $H_{cool}$ induces a very strong anisotropic \cite{ref97} magnetic phase in this nanoparticle system. It was reported earlier that in BiFeO$_3$, the magnetic ordering is of antiferromagnetic type, having a spiral modulated spin structure (SMSS) with an incommensurate long-wavelength period about 62 nm \cite{ref44}, thereby, cancelling any macroscopic magnetization resulting from the spin canting \cite{ref44}. Due to the simultaneous substitution of Gd and Ti in BiFeO$_3$, the SMSS of BiFeO$_3$ is likely suppressed \cite{ref36}. However, the large coercive fields and unsaturated behavior in the hysteresis loop under 30 kOe applied magnetic fields indicate the co-existence of strong-anisotropic ferri/ferromagnetic and antiferromagnetic domains. As a consequence of the coupling between these multiple magnetic domains, it is expected that the system acts as a natural system for generating EB effect in magnetic Bi$_{0.9}$Gd$_{0.1}$Fe$_{0.9}$Ti$_{0.1}$O$_3$ nanoparticles \cite{ref72, ref93}. 

 Although the mechanism for the EB phenomenon of pure nanoparticle is still elusive, however, the physical origin of exchange bias is rather generally accepted to be the exchange coupling between two magnetization components of the individual particle: the uncompensated spins at the surface are ferromagnetic, while the spins in the inner of nanoparticle are antiferromagnetic and referred to as an AFM core surrounded by an FM shell \cite{ref82}. Therefore, the shift in the hysteresis loop occurs due to an interface exchange coupling between the core and shell of the nanoparticle system \cite{ref83}. This core-shell multiferroic nanoparticles are functional nanomaterials and could be a promising candidate for future applications even in several fields of biomedicine \cite{ref11}. It is reported that the interface exchange coupling may provide the nanoparticles with exchange bias properties, which can help in the conversion of electromagnetic energy into heat for magnetic hyperthermia \cite{ref12, ref13}.

\section{Conclusions} \label{II}

In conclusion, tunable exchange bias effect has been observed at temperatures up to 250 K in Bi$_{0.9}$Gd$_{0.1}$Fe$_{0.9}$Ti$_{0.1}$O$_3$ nanoparticles without necessarily cooling the samples through ordering temperature $T_N$. The exchange bias effect depends on cooling magnetic field and the biasing fields are significantly higher at such a high temperature compared than those observed in related materials system at low temperature \cite{ref10,ref26}. Moreover, the large exchange bias field as well as coercive field (up to 1.95 kOe and 5.1 kOe, respectively) make the studied material an interesting starting point for further improving EB materials towards room temperature.The observed shift in the hysteresis loop could speculatively be attributed to an exchange bias effect between permanent antiferromagnetic (and canted) core spins and relative free ferromagnetic surface spins as seen in other nanoparticle systems \cite{ref72, ref93}. The likely presence of both magnetoelectric and interface exchange coupling in this type of multiferroic material system might be worthwhile for their application in novel multifunctional devices. We also anticipate great potential for applications of multiferroic nanoparticles in energy related applications. In the subsequent investigation,  we are interested in solar hydrogen production via water splitting where these nanoparticles will be used as a photocatalyst due to their enhanced photocatalytic activity \cite{ref45} and small band gap. The synthesized photocatalyst will also be deposited as a thin film on a substrate to form a photo-anode (or photoelectrode) for carrying out the water splitting reaction in solution.

\section{Acknowledgements}
The world Academy of Sciences (TWAS), Ref.:14-066 RG/PHYS/AS-I; UNESCO FR: 324028567. This work was also supported by JSPS KAKENHI (No. 26810117) and Nanotechnology Platform Program (Molecule and Material Synthesis) of MEXT, Japan. 
The authors thank to Mr. Motoyasu Fujiwara at the Institute of Molecular Science (IMS) for his assistance during SQUID measurement.

\end{document}